\documentclass{IOS-Book-Article}
\input epsf.sty
\newcommand{\comment}[1]{}

\def \lket {|}
\def \rket {\rangle}
\def \lbra {\langle}
\def \rbra {|}

\newcommand{\ket}[1]{\lket #1\rket}
\newcommand{\bra}[1]{\lbra #1\rbra}

\def\bbbz{Z}
\newtheorem{Theorem}{Theorem}
\newtheorem{Lemma}{Lemma}

\newcommand{\qed}{\nobreak \ifvmode \relax \else
      \ifdim\lastskip<1.5em \hskip-\lastskip
      \hskip1.5em plus0em minus0.5em \fi \nobreak
      \vrule height0.75em width0.5em depth0.25em\fi}

\begin{document}
\begin{frontmatter}                           

\title{Quantum algorithms for formula evaluation}%
\runningtitle{Quantum Algorithms for Formula Evaluation}

\author[A]{\fnms{Andris} \snm{Ambainis}%
\thanks{Faculty of Computing,
University of Latvia, Raina bulv. 19, Riga, LV-1586, 
Latvia, e-mail:{\tt ambainis@lu.lv}.
Supported by European Social Fund project 1DP/1.1.1.2.0/09/APIA/VIAA/044 and Marie Curie 
International Reintegration Grant (IRG)}},

\runningauthor{A. Ambainis}
\address[A]{Faculty of Computing, University of Latvia}

\begin{abstract}
We survey the recent sequence of algorithms for evaluating Boolean formulas
consisting of NAND gates. 
\end{abstract}

\end{frontmatter}

\section{Introduction}

One of two most famous quantum algorithms is Grover's search \cite{Grover}
which can search among $N$ possibilities
in $O(\sqrt{N})$ steps. This provides a quadratic speedup over the naive
classical algorithm for a variety of search problems \cite{Ambainis-survey}.

Grover's algorithm can be re-cast as computing OR of $N$ bits $x_1, \ldots, x_N$,
with $O(\sqrt{N})$ queries to a black box storing 
$x_1, \ldots, x_N$.
A natural generalization of this problem is computing the value 
of an AND-OR formula of $x_1, \ldots, x_N$. 

Grover's algorithm easily generalizes to computing AND-OR formulas of small depth $d$.
Then, $O(\sqrt{N}\log^{d-1} N)$ queries are 
sufficient to evaluate the formula \cite{BCW,Ambainis-var}.
For balanced formulas (with each AND and OR in the formula having the same fan-in),
this can be improved to $O(\sqrt{N})$ \cite{HMW} which is optimal \cite{Ambainis-lb}.

A different case is when, instead of a constant depth, we have a constant fan-in.
This case has been much harder and, until 2007, there has been no 
progress on it at all. If we restrict to binary AND-OR trees, the 
classical complexity of evaluating a full binary AND-OR tree
is $\Theta(N^{.754...})$ \cite{Santha,SW,Snir} and there was no better quantum
algorithm known.

In a breakthrough result, Farhi et al. \cite{FGG} showed that  
the full binary AND-OR tree can be evaluated in $O(\sqrt{N})$ quantum time
in an unconventional continuous-time Hamiltonian query model of \cite{FG,Mochon}.

Several improvements followed soon.
Ambainis et al. \cite{Childs-v1,Ambainis-NAND,Childs-v3,AC+} translated the algorithm of
\cite{FGG} to the conventional discrete time quantum query model and
extending it to evaluating arbitrary Boolean formulas in $O(N^{1/2+o(1)})$ steps. 
Reichardt and \v Spalek \cite{RS,R2009,R2009a} then further extended the algorithm to 
evaluating {\em span programs}, a generalization of Boolean logic formulae. 
This resulted in a surprising result: Reichardt \cite{R2009} showed that
the span-program based approach gave nearly-optimal query algorithms
for any Boolean function. Also using the span program approach, Reichardt \cite{R2009b} 
gave a better formula evaluation algorithm, which can evaluate any Boolean formula
in $O(\sqrt{N} \log N)$ steps (instead of $O(N^{1/2+o(1)})$ in \cite{AC+}).

In this paper, we give a simple description of the basic technical ideas behind
this sequence of quantum algorithms, by describing how the algorithms of 
\cite{FGG,Childs-v1,Ambainis-NAND,Childs-v3,AC+} work for the simplest particular case -
the full binary tree. Besides the two published algorithms \cite{FGG,AC+}, we also describe 
two intermediate versions which appeared in the technical reports \cite{Childs-v1,Ambainis-NAND}.

\section{Technical preliminaries}

\subsection{The problem and motivation}

We consider evaluating a Boolean formula of variables $x_1, \ldots, x_N$
consisting of ANDs and ORs, with each variable occuring exactly once in the formula.
Such a formula can be described by a tree, with variables $x_i$ at the leaves 
and AND/OR gates at the internal nodes. 
This problem has many applications because Boolean formulas can be used to
describe a number of different situations. The most obvious one is 
determining if the input data $x_1, \ldots, x_N$ satisfy certain
constraints that can be expressed by AND/OR gates.

For a less obvious application, we can view
formula evaluation as a black-box model for 
a 2-player game (such as chess) if both
players play their optimal strategies. 
In this case, the game can be represented by a game tree consisting
of possible positions. The leaves of a tree correspond to the possible end
positions of the game. Each of them contains a variable $x_i$, with $x_i=1$
if the $1^{\rm st}$ player wins and $x_i=0$ otherwise. Internal nodes corresponding
to positions which the $1^{\rm st}$ player makes the next move contain a value
that is OR of the values of their children. (The $1^{\rm st}$ player wins if he has a move 
that leads to a position from which he can win.) Internal nodes for 
which the $2^{\rm nd}$ player makes the next move contain a value
that is AND of the values of their children. (The $1^{\rm st}$ player wins if he wins
for any possible move of the $2^{\rm nd}$ player.)

The question is: assuming we have no further information about the game 
beyond the position tree, how many of the variables $x_i$ do we have
to examine to determine whether the $1^{\rm st}$ player has a winning strategy?

\subsection{The model}

By standard rules from Boolean logic (de Morgan's laws), we can replace both AND 
and OR gates by NAND gates. A NAND gate $NAND(y_1, \ldots, y_k)$ outputs 1 
if $AND(y_1, \ldots, y_k)=0$ (i.e., $y_i=0$ for at least one $i\in\{1, \ldots, k\}$)
and 0 otherwise. Then, we have a tree with $x_1, \ldots, x_N$ at the leaves
and NAND gates at the internal vertices. The advantage of this transformation
is that we now have to deal with just one type of logic gates (instead of two - AND 
and OR).

We work in the quantum query model. In the discrete-time version of this model \cite{Ambainis-survey,BWSurvey}, the input bits $x_1, \ldots, x_N$ 
can be accessed by queries $O$ to a  black box.

To define $O$, we represent basis states as $|i, z\rangle$ where
$i\in\{0, 1, \ldots, N\}$. The query transformation $O_x$ 
(where $x=(x_1, \ldots, x_N)$) maps $\ket{0, z}$ to $\ket{0, z}$ and 
$\ket{i, z}$ to $(-1)^{x_i}\ket{i, z}$ for $i\in\{1, ..., N\}$
(i.e., we change phase depending on $x_i$, unless $i=0$ in which case we do
nothing).

Our algorithm can involve queries $O_x$ and arbitrary non-query transformations
that do not depend on $x_1, \ldots, x_N$. The task is to solve a 
computational problem (e.g., to compute a value of a NAND formula)
with as few queries as possible.

In the continuous time Hamiltonian model (first considered by Farhi and Gutman
\cite{FG}), instead of unitary oracle $O_x$, we have a Hamiltonian oracle $H_x$.
We can define $H_x$ as \cite{CG+} \[ H\ket{i} = x_i \ket{i} \]
where $i$ is a register that can hold values $0, 1, \ldots, N$. 
(Similarly to the discrete case, $H_x\ket{0}=\ket{0}$, i.e. we have the option
of not querying any $x_i$.)

We are allowed to combine $H_x$ with an arbitrary time-dependent Hamiltonian $H_0(t)$
that does not depend on $x_1, \ldots, x_N$. The task is to solve the computational
problem by running $H_x$ for as little time as possible.

The continuous time and the discrete query models are roughly equivalent
\cite{Childs-relate,Cleve,CG+}.
A discrete query can be simulated using a Hamiltonian oracle.
And a Hamiltonian query algorithm that uses $T$ queries can be 
transformed into a discrete query algorithm with $O(T \log T/\log \log T)$
queries \cite{CG+}.



\section{Continuous time quantum algorithms}

\subsection{Farhi et al.: quantum walk on an infinite line}

\begin{figure*}
\begin{center}
\epsfxsize=3in
\hspace{0in}
\epsfbox{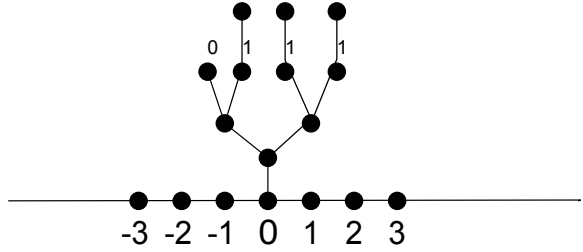}
\caption{An tree augmented by an infinite line and extra edges. The numbers at leaves
show the values of the variables.}
\label{fig:farhi}
\end{center}
\end{figure*}

Assume that we have a formula described by a full binary NAND tree of depth $d$. 
(That is, all variables $x_i$ are at depth $d$. At levels $0, 1, \ldots, d-1$,
we have NAND gates, each of which evaluates NAND of two gates from the next level.)
We augment this tree:
\begin{enumerate}
\item
For each leaf $v$ that contains a variable $x_i=1$, we create a new vertex $v'$, with an edge
$(v, v')$.
\item
We take an infinite line\footnote{A finite segment of the line that is sufficiently long in both directions
can be used as a good approximation to the infinite line. But, for simplicity of the presentation,
we will assume that an infinite line is being used.} of vertices indexed by integers $x$, with a vertex $x$ connected to 
vertices $x-1$ and $x+1$. We connect the root $r$ of our NAND tree to the vertex 0 on this line.
\end{enumerate}
An example of an augmented tree (for depth $d=2$) is shown in Figure \ref{fig:farhi}. 

We now interpret the adjacency matrix of this augmented tree as a Hamiltonian $H$. 
As shown by Farhi et al. \cite{FGG}, if we apply this Hamiltonian $H$ for
time $O(\sqrt{N})$ to an appropriately chosen starting state 
\[ \ket{\psi}=\sum_{i\leq 0} \alpha_i \ket{i} \]
that has non-zero amplitude only at the locations to the left of 0 ($-1, -2, \ldots$), we get
a state $\ket{\psi'}$ with the following properties:
\begin{enumerate}
\item
If the tree evaluates to 1, almost of the state $\ket{\psi'}$ consists of locations to the right of 0;
\item
If the tree evaluates to 0, almost of the state $\ket{\psi'}$ consists of locations to the left of 0;
\end{enumerate}
Thus, by measuring the state $\ket{\psi'}$, we can determine whether the tree evaluates to 0 or 1.
The behaviour of the algorithm can be understood by 
expressing $\ket{\psi'}$ as a superposition 
of the energy eigenstates of the Hamiltonian $H$. Let $\ket{\psi_E}$ be an energy eigenstate
of $H$ with an energy $E=2\theta$. We express
\[ \ket{\psi_E} = \sum_{n\in \bbbz} \alpha_n \ket{n} + \sum_{v\in T} \alpha_v \ket{v} ,\]
with $\ket{n}$ being the vertices on the infinite line ( ``runway") and
$\ket{v}$ being the vertices in the tree.
One can show that the amplitudes of the vertices on the runway are 
\begin{equation}
\label{eq:runway} 
\alpha_n = \left\{ \begin{array}{ll} 
e^{i \theta n} + R(E) e^{-i \theta n} & \mbox{~if $n < 0$} \\
T(E) e^{i \theta n} & \mbox{~if $n \geq 0$} 
\end{array} \right. ,
\end{equation}
where coefficients $R(E)$ and $T(E)$ depend on the energy $E$ and the
structure of the tree (i.e., which leaves of the tree contain $x_i=1$ and, therefore,
have an extra edge attached to them).

$R(E)$ and $T(E)$ are called {\em reflection} and {\em transmission} coefficients
of the tree, by a following physical analogy. We can view tree as an obstacle attached to the 
runway at $n=0$. If we have a particle propagating rightwards from the $n<0$ side,
the particle may either get reflected back to $n<0$ (in which can it starts moving to the
left) or it may pass to $n>0$ and keep moving to the right. 
The reflection and the transmission coefficients describe the amplitudes of
these two possibilities.
We have 

\begin{Theorem}
\cite{FGG}
\begin{enumerate}
\item
If $F=0$, then $T(0)=0$ and $R(0)=-1$.
\item
If $F=1$, then $T(0)=1$ and $R(0)=0$.
\end{enumerate}
\end{Theorem}

Thus, if $F=0$, we have an eigenstate $\ket{\psi_0}$ which has zero amplitudes 
for $n\geq 0$. This eigenstate is
\begin{equation}
\label{eq:psi0} 
\ket{\psi_0} = \sum_{k \geq 0} (\ket{-4k} - \ket{-4k-2})  
+ \sum_{v\in T} \alpha_v \ket{v} 
\end{equation}
for some amplitudes $\alpha_v$.
If we start in $\ket{\psi_0}$\footnote{More precisely,
we start in a slightly different state 
$\ket{\psi_{start}}=\sum_{k \geq 0} (\ket{-4k} - \ket{-4k-2})$
instead of $\ket{\psi_0}$, since we do not know the amplitudes $\alpha_v$.
It can be shown that $\ket{\psi_{start}}$ is a sufficiently good 
approximation of $\ket{\psi_0}$.}
and apply $H$, the state $\ket{\psi_0}$ stays unchanged.
In contrast, if $F=1$, the same state $\ket{\psi_0}$ is not an eigenstate
and applying $H$ for sufficiently long time leads to nonzero amplitudes for $n\geq 0$.
Thus, we can distinguish $F=0$ and $F=1$ by preparing $\ket{\psi_0}$,
applying $H$ and measuring the state. If we find $\ket{n}$, $n\geq 0$,
we know that $F=1$.

A slight complication is that the states $\ket{\psi_0}$ and $\ket{\psi_{start}}$ 
has equal amplitudes on infinitely many states $\ket{-4k}$ 
and $\ket{-4k-2}$ and, thus has an infinite norm. 
This can be avoided by using
\[ \ket{\psi'_{start}} = \frac{1}{\sqrt{2L}} \sum_{k=0}^{L-1} 
(\ket{-4k} - \ket{-4k-2}) .\]
This state turns out to be a sufficiently good approximation of 
$\ket{\psi_{start}}$. 

\subsection{Childs et al.: finite segment}
\label{sec:childs1}

A modification of FGG algorithm was proposed by Childs et al. \cite{Childs-v1}.

\begin{figure*}
\begin{center}
\epsfxsize=3in
\hspace{0in}
\epsfbox{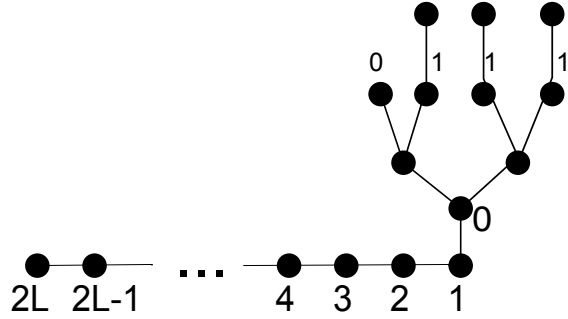}
\caption{A tree augmented by a finite tail and extra edges.}
\label{fig:childs}
\end{center}
\end{figure*}

We can, instead of an infinite line, attach a finite segment of length $2L$ 
to the root of the tree (see figure \ref{fig:childs}).
The starting state is similar to the infinite line algorithm:
\[ \ket{\psi'_{start}} = \frac{1}{\sqrt{L+1}} \sum_{k=0}^{L} 
(-1)^k \ket{2k} \]
where $L=O(\sqrt{N})$. (Here, $\ket{0}$ denotes the root of the tree
and $\ket{1}, \ldots, \ket{2L}$ denote the vertices in the tail.)

In this case, we have the following behaviour. If $F=0$, 
the state $\ket{\psi'_{start}}$ remains almost unchanged by $H$.
If $F=1$, then, after $O(\sqrt{N})$ steps, the state becomes 
sufficiently different from $\ket{\psi'_{start}}$.
More formally:

\begin{Lemma}
\begin{enumerate}
\item
If $F=0$, then there exists $\ket{\psi}$ such that $H\ket{\psi}=0$
and $\lbra \psi \ket{\psi'_{start}}\geq 1-\epsilon$.
\item
If $F=1$, then, for any eigenstate $\ket{\psi}$ of Hamiltonian $H$, either 
the corresponding eigenvalue $\lambda$ satisfies $|\lambda| =
\Omega(\frac{1}{\sqrt{N}})$ or $\ket{\psi} \perp \ket{\psi_{start}}$.
\end{enumerate}
\end{Lemma}

Because of that, we can distinguish between the two cases by running the eigenvalue
estimation \cite{Eigenvalue,Eigenvalue1} on state $\ket{\psi_{start}}$, with precision
$\frac{1}{\sqrt{N}}$. If $F=0$, we get the answer $\lambda=1$ with a high
probability 1. If $F=1$, we get a value $\lambda$ with 
$|\lambda| \geq \frac{1}{\sqrt{N}}$.
To estimate the eigenvalue with precision $\frac{1}{\sqrt{N}}$,
it is sufficient and necessary to run the Hamiltonian $H$ for time $O(\sqrt{N})$.

The computer experiments suggest that we can also distinguish between the two cases
simply by running the Hamiltonian for time $O(\sqrt{N})$ and measuring the final state.
If $F=0$, we find one of basis states $\ket{2k}$ with a high probability.
If $F=1$, we find one of basis states $\ket{v}$ in the tree.

\section{Discrete time algorithms}

There are two ways to transform the above algorithm into a discrete time quantum algorithm,
discovered independently
by Ambainis \cite{Ambainis-NAND} and Childs et al. \cite{Childs-v3}.


\subsection{Discrete time algorithm by eigenspace decomposition}
\label{sec:ambainis}

The basic idea behind the algorithm of \cite{Ambainis-NAND} is as follows.
We can decompose the continuous time Hamiltonian $H$ as $H=H_{tree}+H_{input}$
where $H_{tree}$ is the part of $H$ that is independent of $x_1, \ldots, x_N$
and $H_{input}$ consists of extra edges that are added to the tree
if $x_i=1$. To obtain similar behaviour in discrete time, we define 
two unitary transformations $U_{tree}$ and $U_{input}$ that correspond 
to $H_{tree}$ and $H_{input}$ and then consider $U=U_{input} U_{tree}$.

In the continuous case we had $F=0$ iff there exists 
$\ket{\psi}\approx \ket{\psi_{start}}$ with
\[ H\ket{\psi} = H_{tree} \ket{\psi} + H_{input} \ket{\psi} = 0 .\]
In the discrete time, this corresponds to
$ U \ket{\psi} = U_{tree} U_{input} \ket{\psi} = \ket{\psi} $.

To define $U_{tree}$ and $U_{input}$, we consider a tree that is augmented by a
finite tail of length $2L$ (as in the previous algorithm) but with no extra edges
at the leaves. $U_{tree}$ is defined by
\[ 
U_{tree} \ket{\psi} = \left\{ \begin{array}{ll} 
\ket{\psi} & \mbox{~if $H_{tree}\ket{\psi} = 0$} \\
-\ket{\psi} & \mbox{~if $H_{tree}\ket{\psi} = \lambda \ket{\psi}$, $\lambda\neq 0$} \\
\end{array} \right.
\]
$U_{input}$ is defined as follows. $U_{input}\ket{v}=-\ket{v}$ if $v$ is a leaf containing
$i:x_i=1$ and $U_{input}\ket{v}=\ket{v}$ if $v$ is a leaf containing
$i:x_i=0$ or if $v$ is an internal vertex.

\begin{figure*}
\begin{center}
\epsfxsize=3in
\hspace{0in}
\epsfbox{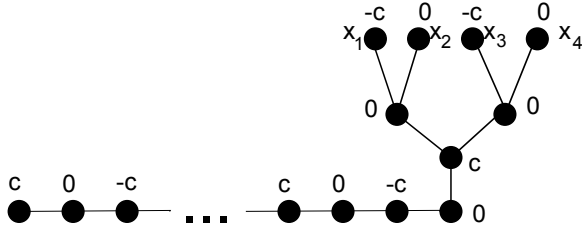}
\caption{A 0-eigenstate of $H$. $c$ is a number that is chosen so that
the norm of the state is 1.}
\label{fig:4}
\end{center}
\end{figure*}

If $F=0$, there is a state $\ket{\psi_0}$ satisfying $U_{tree}\ket{\psi_0}=U_{input}\ket{\psi_0}= \ket{\psi_0}$ and 
$\ket{\psi_0}\approx \ket{\psi_{start}}$.
In figure \ref{fig:4}, we show this state for a NAND tree of depth 2,
with particular values of variables ($x_1=x_3=1$, $x_2=x_4=0$.)

If $F=1$, there is no such state. To see that, we first notice that $U_{tree}$
and $U_{input}$ are both reflections. Therefore, 
the spectrum of $U_{input}U_{tree}$ can be analyzed within
the "two reflections" framework of Aharonov \cite{Aha99}
and Szegedy \cite{Szegedy}. We show

\begin{Lemma}
\cite{Ambainis-NAND}
Let $H_1$ be the subspace spanned by $\ket{v}$ for leaves $v$ containing variables $x_i=1$
and let $P_1$ be the projection to $H_1$.
Assume that, for any 1-eigenvector $\ket{\psi}$: $U_{tree}\ket{\psi}=\ket{\psi}$ which
is not orthogonal to $\ket{\psi_{start}}$, we have
\[ \| P_1 \ket{\psi} \| \geq \epsilon \|\psi\| .\]
Then, any eigenvector $\ket{\psi}$: $U\ket{\psi} = \lambda \ket{\psi}$ which is  not
orthogonal to $\ket{\psi_{start}}$ must have $|\lambda-1| \geq \epsilon$.
\end{Lemma}

Therefore, to show the desired bound on the eigenvalues of $U$ ($\lambda=e^{i\theta}$,
$\theta=\Omega(1/\epsilon)$), it suffices to lower-bound the projections of eigenvectors
of $U_{tree}$ to $H_1$. This is done by analyzing the amplitudes of $\ket{\psi}$
at vertices of the tree that evaluate to 1 (either internal vertices which contain
a NAND gate whose output is 1 or leaves that contain $x_i=1$). We start by analyzing the amplitude of the root (which evaluates to 1 because $F=1$) and then move up the tree. See 
\cite{Ambainis-NAND} for details.

\subsection{Discrete time algorithm via Szegedy quantization}
\label{sec:childs2}

An alternative approach is to construct a quantum algorithm based on discrete time 
quantum walk. As in the previous section, we consider a NAND tree augmented by a 
finite tail of even length $L=O(\sqrt{N})$. We consider a coined quantum walk 
on this tree defined in the natural way \cite{Ambainis-ICQI}.
The coined walk has the state space consisting of states $\ket{v, d}$
where $v$ is a vertex of the augmented tree, $d\in\{ down, left, right \}$ and:
\begin{itemize}
\item 
If $v=L$ (i.e., $v$ is the end of the tail), then $d=left$;
\item
If $v\in\{1, \ldots, L-1\}$ (i.e., $v$ is in the tail but is not the end of the tail),
then $d\in\{left, right\}$);
\item
If $v$ is a non-leaf vertex in the tree, then $d\in\{down, left, right\}$;
\item
If $v$ is a leaf, then $d=down$. 
\end{itemize}
Each such state $\ket{v, d}$ can be associated with one of edges incident to $v$,
in a natural way. If $v$ is a vertex in the tail, then states 
$\ket{v, left}$ and $\ket{v, right}$ correspond to edges $(v, v+1)$ and $(v, v-1)$,
respectively. If $v$ is a vertex in the tree, $\ket{v, down}$ corresponds to
the edge $(v, p)$ where $p$ is $v$'s parent and $\ket{v, left}$ and $\ket{v, right}$
correspond to edges $(v, l)$ and $(v, r)$ where $l$ and $r$ are $v$'s left
and right child.

A discrete-time quantum walk (also called coined quantum walk) consists of two steps:
a coin flip $C$ and a shift operator $S$.
The coin flip $C$ is defined as follows: for each $v$, we apply a transformation
$C_v$ on the subspace $H_v$ spanned by states $\ket{v, d}$.
More specifically:
\begin{itemize}
\item
If $v=L$, $C_v=I$.
\item
If $v$ is a leaf, $C_v=(-1)^{x_i} I$ where $x_i$ is the variable at the leaf $v$.
\item
Otherwise, $C_v=2\ket{\psi}\bra{\psi}-I$ where $\ket{\psi}$ is the uniform
superposition over all possible $\ket{d}$. 
(For states $v\in\{1, \ldots, L-1\}$ in the tail, $\ket{\psi}=\frac{1}{\sqrt{2}}
(\ket{left}+\ket{right})$. For states $v$ in the tree,
$\ket{\psi}=\frac{1}{\sqrt{3}} (\ket{left}+\ket{right}+\ket{down})$.)
\end{itemize} 
A shift $S$ is just the transformation that, for every edge $(v, v')$ in the
augmented tree, swaps the two basis states $\ket{v, d}$ and $\ket{v', d'}$
associated with the edge $(v, v')$.

There is a close correspondence between the eigenvalues of Hamiltonian $H$ 
(described in section \ref{sec:childs1}) and the eigenvalues of 
the unitary transformation $SC$. Namely, let the starting state be
\[ \ket{\psi_{start}} = \frac{1}{\sqrt{4L}} \sum_{k=0}^{L-1} 
(-1)^k ( \ket{2k, right}-i \ket{2k+1, left} + \]
\begin{equation}
\label{eq:long} i \ket{2k+1, right} - \ket{2k+2, left} ) .
\end{equation}
Then, we have:
\begin{enumerate}
\item
If $F=0$, there exists $\ket{\psi}$ such that $\|\ket{\psi}-\ket{\psi_{start}}\|
\leq \epsilon$ and $SC \ket{\psi} = i \ket{\psi}$.
\item
If $F=1$, then, for any eigenstate $\ket{\psi}$ (with $CS\ket{\psi}=\lambda \ket{\psi}$),
either $\ket{\psi} \perp \ket{\psi_{start}}$ or $Re\lambda = c/\sqrt{N}$ for some
constant $c>0$.
\end{enumerate}
Running the eigenvalue estimation with $\ket{\psi_{start}}$ as the starting state
and accuracy $c/2\sqrt{N}$ allows us to distinguish between the two cases. 

The finite tail can be shortened from $L$ vertices to just 2. 
Namely \cite{Childs-v3}, we can attach a tail of length 2 consisting of two
vertices 1 and 2. For the tree vertices, we define quantum walk in the same way as before.
For the vertex 1, we define $C_1 = 2\ket{\psi}\bra{\psi}-I$ where
\[ \ket{\psi}= \frac{1}{\sqrt[4]{N}} \ket{right} + 
\sqrt{1-\frac{1}{\sqrt{N}}} \ket{left} .\]
For the vertex 2, we define $C_2=I$.

Then, taking the starting state $\ket{\psi_{start}}=\ket{2}$ gives us the same
behaviour as we had for the starting state given by (\ref{eq:long}) for the walk
with tail of length $L$.

\section{Further developments}

All of those algorithms can be generalized to evaluating NAND trees of arbitrary 
structure. The key idea is to use a {\em weighted adjacency matrix} as the Hamiltonian. 
That is, for every edge $(u, v)$ in the tree, we define a weight $w_{uv}$ that depends on
the number of leaves in the subtree above $u$ and the number of leaves in the subtree 
above $v$. We then take the matrix $H$ defined by $H_{uv}=0$ if $(u, v)$ is not an edge 
and $H_{uv}=w_{uv}$ if $(u, v)$ is an edge. 

If a formula $F$ is of depth $d$, we can choose weights $w_{uv}$ so that a similar 
quantum algorithm (with the weighted $H$ as the Hamiltonian) evaluates 
$F$ with $O(\sqrt{N d})$ queries 
\cite{Childs-v3,Ambainis-NAND,AC+}. For formulas $F$ of large depth $d$, one
should first reduce their depth using a following result of Bshouty, Cleve and Eberly \cite{BCE}:

\begin{Theorem}
\cite{BCE}
For any NAND formula $F$ of size $N$, 
there exists a NAND formula $F'$ of size $N'=O(N^{1+O(\frac{1}{\sqrt{\log N}})})$
and depth $d'=O(N^{O(\frac{1}{\sqrt{\log N}})})$ such that $F'=F$.
\end{Theorem}

By first applying this theorem to reduce the depth
and then using the $O(\sqrt{N'd'})$ query algorithm to evaluate the resulting $T'$,
we can evaluate any Boolean formula $T$ with $O(N^{1/2+O(1/\sqrt{\log N})})$
queries \cite{Childs-v3,Ambainis-NAND,AC+}.

The NAND tree evaluation algorithms have been generalized by Reichardt and 
\v Spalek \cite{RS} and Reichardt \cite{R2009,R2009a,R2009b}. 
They have discovered that similar ideas can be used to evaluate 
{\em span programs} (an algebraic computation model which generalizes
Boolean logic formulas) with the number of queries being square
root of the {\em witness size} of the span program \cite{RS}.

The span program framework is very powerful. It has been used to design better
quantum algorithms for many specific Boolean formulas (by designing span programs
for them) \cite{RS}. Moreover, Reichardt \cite{R2009,R2010} has shown that 
span programs are nearly optimal for any Boolean function.
That is, if a Boolean function can be evaluated with $t$ queries by some
quantum algorithm, then:
\begin{itemize}
\item
It can be evaluated by a span-program based quantum algorithm, 
using a generalization of the algorithm from section \ref{sec:childs2}
with $O(t \frac{\log t}{\log \log t})$ queries \cite{R2009};
\item
It can be evaluated by a span-program based quantum algorithm, 
using a generalization of the algorithm from section \ref{sec:childs2}
with $O(t)$ queries \cite{R2010}.
\end{itemize}
Span programs also can be used to 
evaluate any NAND formula (of any depth) with $O(\sqrt{N} \log N)$
queries \cite{R2009b}.

\end{document}